\documentclass[12pt,titlepage]{article}
\usepackage[letterpaper,top=1in,bottom=1in,left=1.25in,right=1.25in]{geometry}
\usepackage{graphicx,cite, url}

\begin{document}
\title{Computational illumination for high-speed \emph{in vitro} Fourier ptychographic microscopy}

\author{Lei Tian$^{1,\star,*}$, Ziji Liu$^{1,2,\star}$, Li-Hao Yeh$^1$, Michael Chen$^1$, Jingshan Zhong$^1$, \\Laura Waller$^1$\\
\\
\multicolumn{1}{p{\textwidth}}{\centering\emph{\normalsize 1. Department of Electrical Engineering and Computer Sciences, University of California Berkeley, Berkeley, 94720, USA\\
2. State Key Laboratory of Electronic Thin Films and Integrated Devices, School of Optoelectronic Information, University of Electronic Science and Technology of China (UESTC), Chengdu, 610054, China\\
$^{\star}$ These authors contributed equally to this work\\
$^{*}$ lei\_tian@alum.mit.edu
}}}

\maketitle

\begin{abstract}
We demonstrate a new computational illumination technique that achieves large space-bandwidth-\emph{time} product, for quantitative phase imaging of unstained live samples \emph{in vitro}. Microscope lenses can have either large field of view (FOV) or high resolution, not both. Fourier ptychographic microscopy (FPM) is a new computational imaging technique that circumvents this limit by fusing information from multiple images taken with different illumination angles. The result is a gigapixel-scale image having both wide FOV and high resolution, i.e. large space-bandwidth product (SBP). FPM has enormous potential for revolutionizing microscopy and has already found application in digital pathology. However, it suffers from long acquisition times (on the order of minutes), limiting throughput. Faster capture times would not only improve imaging speed, but also allow studies of live samples, where motion artifacts degrade results. In contrast to fixed (e.g. pathology) slides, live samples are continuously evolving at various spatial and temporal scales. Here, we present a new source coding scheme, along with real-time hardware control, to achieve 0.8 NA resolution across a 4$\times$ FOV with sub-second capture times. We propose an improved algorithm and new initialization scheme, which allow robust phase reconstruction over long time-lapse experiments. We present the first FPM results for both growing and confluent \emph{in vitro} cell cultures, capturing videos of subcellular dynamical phenomena in popular cell lines undergoing division and migration. Our method opens up FPM to applications with live samples, for observing rare events in both space and time.
\end{abstract}

\section{Introduction}
\emph{In vitro} microscopy is crucial for studying physiological phenomena in cells. For many applications, such as drug discovery~\cite{Lang2006}, cancer cell biology~\cite{Boyd1995} and stem cell research~\cite{Costa2011}, the goal is to identify and isolate events of interest. Often, these events are rare, so automated high throughput imaging is needed in order to provide statistically and biologically meaningful analysis~\cite{Rimon2011}. Thus, an ideal technique should be able to image and analyze thousands of cells simultaneously across a wide field of view (FOV). To observe dynamical processes across various spatial and temporal scales, both high spatial and high temporal resolution are required. Existing high throughput imaging techniques~\cite{Zheng2011a, Greenbaum2012, Zheng2013, Greenbaum2014, Luo2015, Goda2009} cannot meet the space-bandwidth-\emph{time} product (SBP-T) required for widefield \emph{in vitro} applications. Here, we introduce a new computational microscopy technique with both high spatial and temporal resolution over a wide FOV. 

\begin{figure}[t]
\centerline{\includegraphics[width=0.75\textwidth]{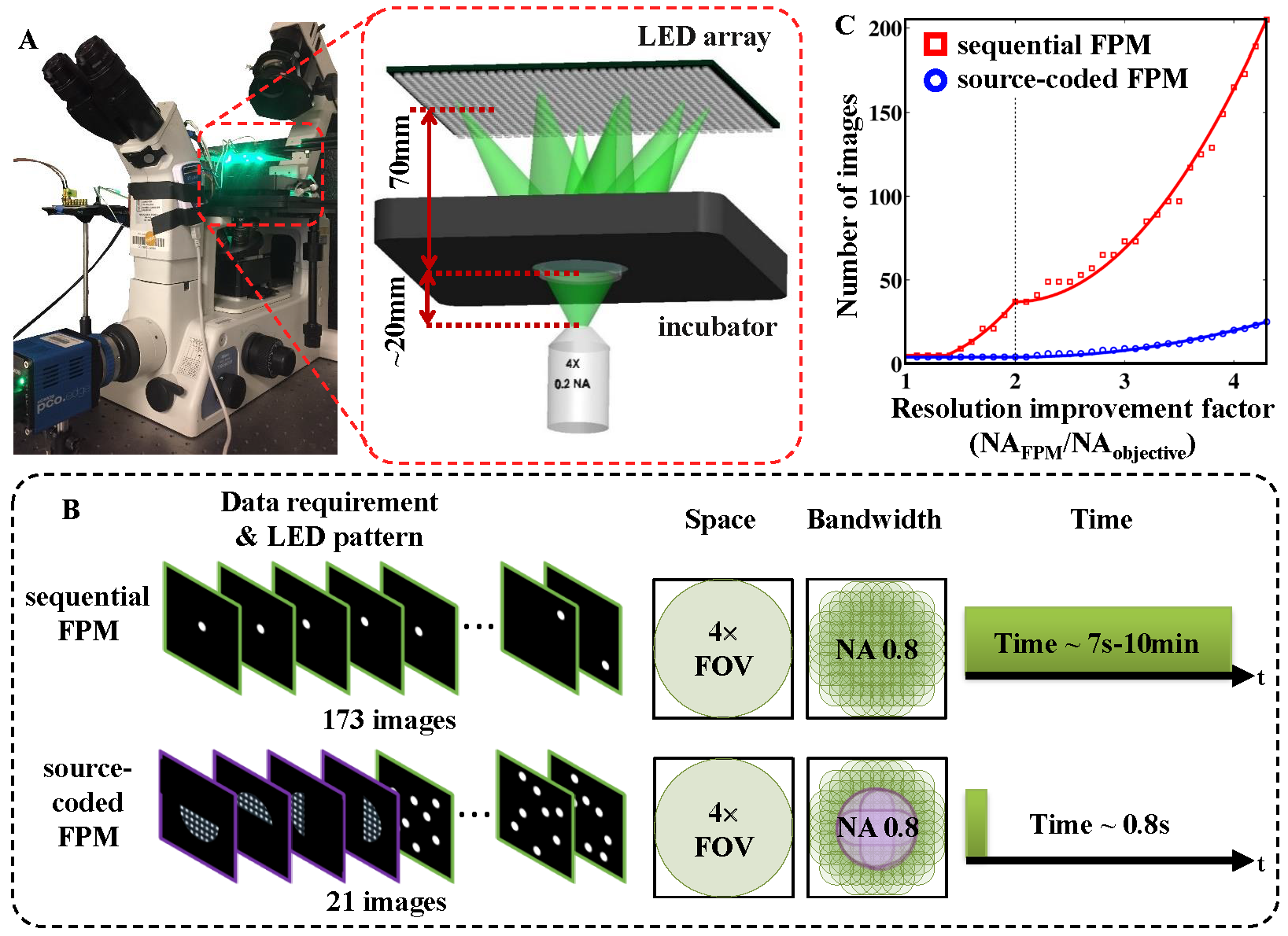}}
\caption{Source-coded Fourier ptychographic microscopy (FPM) captures large space-bandwidth product (SBP) images in under 1 second. (A) Experimental setup is a microscope with an LED array source and a wide FOV 4$\times$ (0.2 NA) objective. Multiple images are captured with coded illumination in order to reconstruct higher resolution (up to 0.8 NA). (B) Comparison of illumination schemes in terms of space, bandwidth and acquisition time. Sequential FPM scans through each LED, achieving large SBP at a cost of speed. Our source-coded FPM implements hybrid patterning to achieve the same SBP with sub-second acquisition time. (C) The number of images required for source-coded FPM (blue) grows more than 8$\times$ slower than for sequential FPM (red) as the final resolution increases. (Solid lines: theoretical, points: our LED array). \label{setup}}
\end{figure}

Our method is an extension of Fourier ptychographic microscopy (FPM)~\cite{Zheng2013}, which overcomes the physical space-bandwidth product (SBP) limit. Instead of choosing between large FOV \emph{or} high resolution, FPM achieves both by trading acquisition speed. Illumination angles are scanned sequentially with a programmable LED array source (Fig.~\ref{setup}), while taking an image at each angle. Tilted illumination samples different regions of Fourier space, as in synthetic aperture~\cite{Gutzler2010, Tippie2011} and structured illumination~\cite{Gustafsson2000, Wicker2014} imaging. Though the spatial resolution of each measurement is low, the images collected with high illumination angles (dark field) contain sub-resolution information. The reconstructed image achieves resolution beyond the diffraction limit of the objective - the sum of the objective and illumination NAs. Distinct from synthetic aperture, FPM does not measure phase directly at each angle, but instead uses nonlinear optimization algorithms~\cite{Zheng2013, Ou:14, Tian2014, Tian2015} similar to translational diversity~\cite{rodenburg2004phase, guizar2008phase} and ptychography~\cite{maiden2009improved, thibault2009probe}. Conveniently, the LED array coded source used here is implemented as a simple and inexpensive hardware modification to an existing microscope. Our proposed method uses efficient source coding schemes to reduce the capture time by several orders of magnitude.

The standard approach for large SBP microscopy is slide scanning, in which the sample is mechanically moved around while imaging with a high resolution objective to build up a large FOV. Scanning limits throughput~\cite{Orth2012} and is unsuitable for \emph{in vitro} imaging of dynamic events. Instead of starting with high resolution and stitching together a large FOV, FPM starts with a large FOV and stitches together images to recover high resolution. A major advantage is its ability to capture the required set of images with no moving parts, by varying illumination angles. In addition, scanners use high magnification objectives with short depth of field (DOF), thus requiring extensive auto-focusing mechanisms. In contrast, FPM provides robustness to focus errors because its DOF is longer than that which would be provided by a high-magnification objective of equivalent NA~\cite{Tian2015}. Further, out of focus images can be digitally refocused~\cite{Zheng2013}.   

The main limitation of FPM for real-time \emph{in vitro} applications is long acquisition times and large datasets. Not only are a large number of images ($\sim$200) captured for each reconstruction, but also long exposure times are needed for the dark field images. To shorten exposure times, we first built a custom hardware setup consisting of high-brightness LEDs and fast control circuitry (see Methods~\ref{method-setup}). This allows us to reduce total acquisition time by $\sim50\times$, enabling a full 173 LED scan to capture 0.96 gigapixels of data  in under 7 seconds. While this speed is already suitable for many \emph{in vitro} applications (\emph{e.g.} cell division processes), fast subcellular dynamics (\emph{e.g.} vesicle tracking) require sub-second capture to avoid severe motion blur. Since the camera's data transfer rate is the limiting factor, large SBP images cannot be captured in less than 1 second unless we can reduce the data requirements. Furthermore, for studies requiring time-series measurements, the large data requirement ($\sim$1 gigapixel per dataset) poses severe burdens on both storage and processing.

In order to reconstruct large SBP from fewer images, we need to fundamentally change our capture strategy. To do this, we eliminate redundancy in the data by designing new source coding schemes. The redundancy arises because of a $60\%$ overlap requirement in Fourier space for neighboring LEDs~\cite{bunk2008influence,Zheng2013}. This means that we must capture $\sim$10$\times$ more data than we reconstruct if using `sequential' FPM. Our angle-multiplexing scheme instead turns on multiple LEDs simultaneously for each measurement, allowing better coverage of Fourier space with each image. Without eliminating the overlap, we therefore fill in Fourier space faster. Previous work employed a random coding strategy across both brightfield and dark field regions~\cite{Tian2014}.  One problem with this scheme is that the brightfield images and dark field images have large differences in intensity ($\sim$10-100$\times$). Considering Poisson noise statistics, this means that images with mixed brightfield and dark field components may suffer from the dark field signal being overwhelmed by the brightfield noise. As a result, we significantly improve our multiplexing results here by separating brightfield from dark field LEDs. Furthermore, we note that asymmetric illumination-based differential phase contrast (DPC)~\cite{mehta2009quantitative, Tian2015a} provides a means for recovering quantitative phase and intensity images out to 2$\times$ the objective NA with only 4 images. Hence, there is no need to individually scan the brightfield LEDs. Our new method, termed source-coded FPM, uses a hybrid illumination scheme: it first captures 4 DPC images (top, bottom, left, right half-circles) to cover the brightfield LEDs, then uses random multiplexing with 8 LEDs to fill in the dark field Fourier space region (Fig.~\ref{setup}B).

Source-coded FPM approaches the theoretical limit for \mbox{SBP-T}~\cite{Lukosz1966, Lukosz1967}, set by the camera's data transfer rate. We achieve \mbox{SBP-T} of 46~megapixels/second, calculated according to~\cite{Zheng2015} (4$\times$ FOV and 0.8 NA captured in 0.8 seconds). Though the data rate of our camera is larger (138 megapixels/second), we leave some redundancy in order to ensure robust algorithm convergence. Reconstruction algorithms that explicitly take \emph{a priori} information into account, such as sparsity based methods~\cite{candes2008introduction, shechtman2013gespar}, could further improve capture speed; here, for generality, we choose not to make any assumptions on the sample. Both sequential FPM and source-coded FPM require the number of images in the dataset to grow quadratically with improved final resolution. This is due to the coverage area in Fourier space increasing in proportion to the square of the final NA. However, our source-coded method has a slower growth rate (Fig.~\ref{setup}C) and stays flat for less than 2$\times$ resolution improvement. Conveniently, these techniques can flexibly trade off FOV, resolution and acquisition time by choosing the illumination angle range. 

For \emph{in vitro} applications, stain-free and label-free is particularly attractive because it is non-invasive and non-toxic. FPM provides both intensity and phase~\cite{ou2013quantitative}, which contain morphological and cell mass~\cite{Popescu2008} information that can be used for quantitative study~\cite{Cohen2010,Mir2011}. In order to achieve accurate, high-quality results for unstained samples, we needed to make several modifications to the FPM algorithm. It is well known that non-convex problems such as phase retrieval will often get stuck in local minima~\cite{Fienup1982,Fienup1986,Yang2011}; the best way to avoid this is to provide a good initial guess~\cite{Fienup1986,Greenbaum2014,Candes2014}. Previous work in FPM uses a low-resolution intensity image as the initial guess~\cite{Zheng2013, Ou:14, Tian2014, ou2013quantitative}. Stained samples with strong intensity variations thus reconstruct successfully, since the intensity-only initialization is close to the actual solution. However, unstained samples are phase objects and so the intensity-only initial guess does not provide a good starting point. Furthermore, FPM does not measure phase directly at each angle, and the phase contrast provided by the asymmetric illumination results in uneven sensitivity to phase at different spatial frequencies. A detailed phase transfer function analysis~\cite{Tian2015a} based on the weak-object (Born) approximation~\cite{Rose1977a, Hamilton1984a} shows that low-frequency phase information is poorly captured, since it results only from illumination angles that are close to the objective NA. Thus, low spatial frequency phase information is more difficult to reconstruct than high spatial frequency phase information, contrary to the situation for intensity reconstructions. To improve our reconstruction, we use a linearly approximated phase solution based on DPC deconvolution~\cite{Tian2015a} as a close initial guess for spatial frequencies within the 2 NA bandwidth. We then run a nonlinear optimization algorithm to solve the full phase problem (see Methods~\ref{method-algorithm}), resulting in high-quality phase reconstructions with high resolution (Fig.~\ref{compare}A) and good low spatial frequency phase recovery (Fig.~\ref{compare}B). 

We demonstrate our new source-coded FPM by reconstructing large SBP videos of popular cell types \emph{in vitro} on a petri dish, for both growing and confluent samples. We observe sub-cellular and collective cell dynamics happening on different spatial and temporal scales, allowing us to observe rare events in both space and time, due to the large space-bandwidth-\emph{time} product and flexible tradeoff of time and SBP.

\section{Results}

\subsection{Validation with stained and unstained samples}

\begin{figure}[!th]
  \centerline{\includegraphics[width=0.99\textwidth]{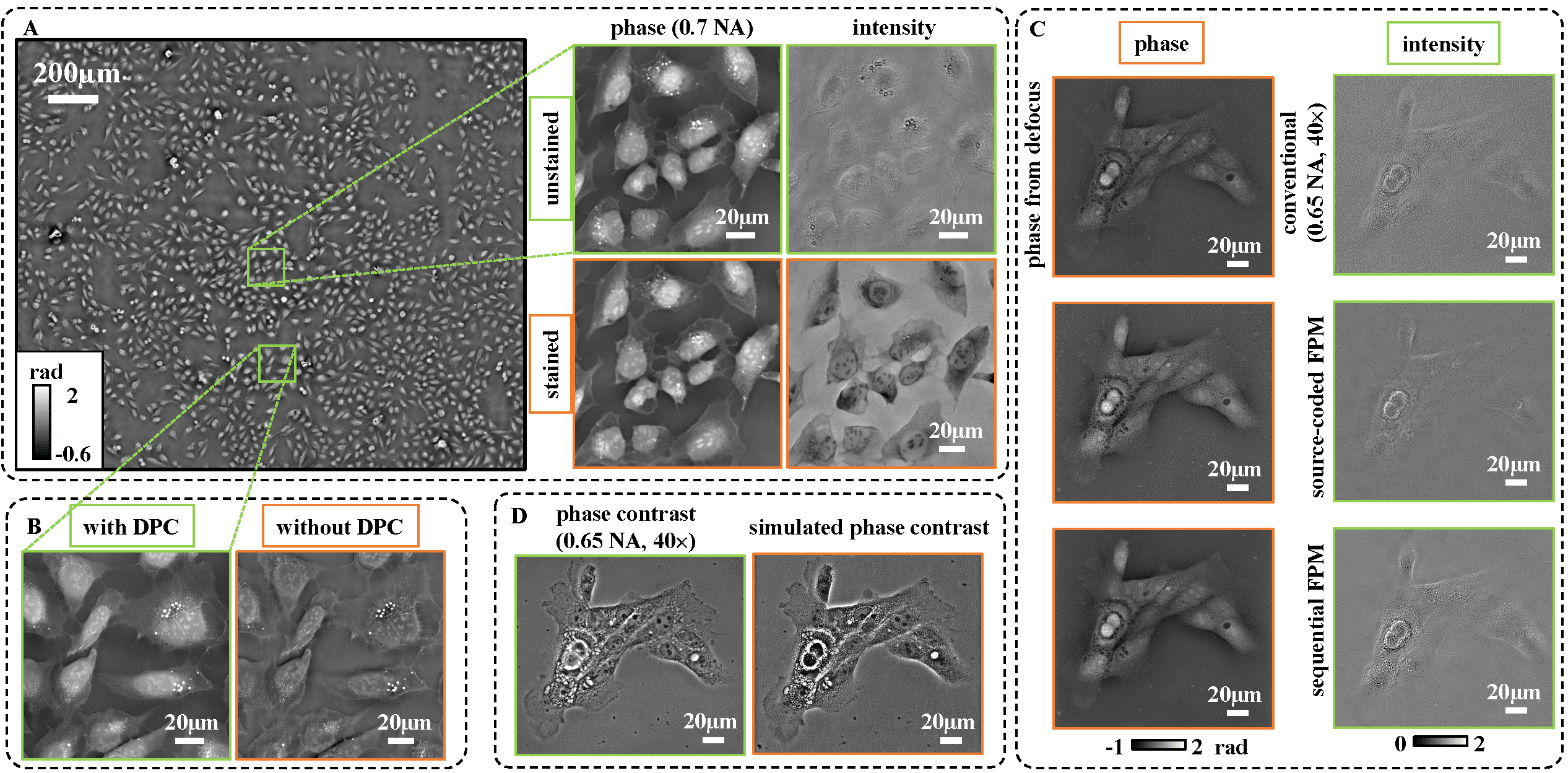}}
\caption{Large SBP reconstructions of quantitative phase and intensity. (A) Phase reconstruction across the full FOV of a 4$\times$ objective with 0.7 NA resolution (sample: U2OS). A zoom-in is shown at right, with comparison to reconstructions of the same sample before and after staining. 
(B) Our improved FPM algorithm provides better reconstruction of low-frequency phase information. A zoom-in region shows comparisons between phase reconstructions with and without our differential phase contrast (DPC) initialization scheme. (C) To validate our source-coded FPM results, we compare to images captured with a 40$\times$ objective having high resolution (0.65 NA) but small FOV (sample: MCF10A), as well as sequential FPM. (D) We simulate a phase contrast image and compare to one captured by a high-resolution objective (0.65 NA, 40$\times$).\label{compare}}
\end{figure}

A major advantage of quantitative phase imaging is that it can visualize transparent samples in a label-free, non-invasive way. Figure~\ref{compare}A compares our reconstructions before and after staining, for the same fixed human osteosarcoma epithelial (U2OS) sample. With staining, the intensity image clearly displays detailed subcellular features. Stained samples also display strong phase effects, proportional to the local shape and density of the sample. Without staining, the intensity image contains very little contrast and is nearly invisible; however, the phase result clearly captures the sub-cellular features. Due to the strong similarity between stained intensity and unstained phase, it follows that quantitative phase may provide a valid alternative to staining. 

To demonstrate the importance of using a good initial guess to initialize the phase recovery for unstained samples, we compare the FPM results both with and without our DPC initialization scheme (Fig.~\ref{compare}B). Both achieve the same 0.7 NA resolution, with high spatial frequency features (e.g. nucleus and filopodia) being clearly reconstructed, as expected. However, without DPC initialization, low frequency components of the phase are not well recovered, resulting in a high-pass filtering effect on the reconstructed phase, much like Zernike phase contrast. With DPC initialization, low frequencies, which describe the overall height and shape of the cells, are correctly recovered.  

Next, we verify the accuracy of the recovered phase values by comparing to images captured directly with a higher resolution objective (40$\times$ 0.65 NA). The  reconstruction resolution for our method matches its expected value (0.7~NA), as seen in Fig.~\ref{compare}C, which shows images of fixed human mammary epithelial (MCF10A) cells. To validate our quantitative phase result, we compare to that recovered from a through-focus intensity stack captured with the high-resolution objective, then input into a phase retrieval algorithm~\cite{Jingshan14GPTIE} (Fig.~\ref{compare}C and Methods~\ref{method-throughfocus}). Quantitative phase can further be used to simulate popular phase contrast modes such as DIC and Zernike phase contrast (PhC)~\cite{JMI:JMI1027}. In Fig.~\ref{compare}D, we compare an actual PhC image to that simulated from our method's reconstructed result (see Methods~\ref{method-PhC}). Since the PhC objective uses annular illumination (Ph2, 0.25~NA) to achieve resolution corresponding to 0.9 NA, it should have slightly better resolution than our reconstruction. Note that the simulated PhC image is effectively high-pass filtered phase information, so small details appear with better contrast.

\subsection{Fast sequential FPM video of HeLa cells dividing \emph{in vitro}}

\begin{figure}[!t]
\centerline{\includegraphics[width=\textwidth]{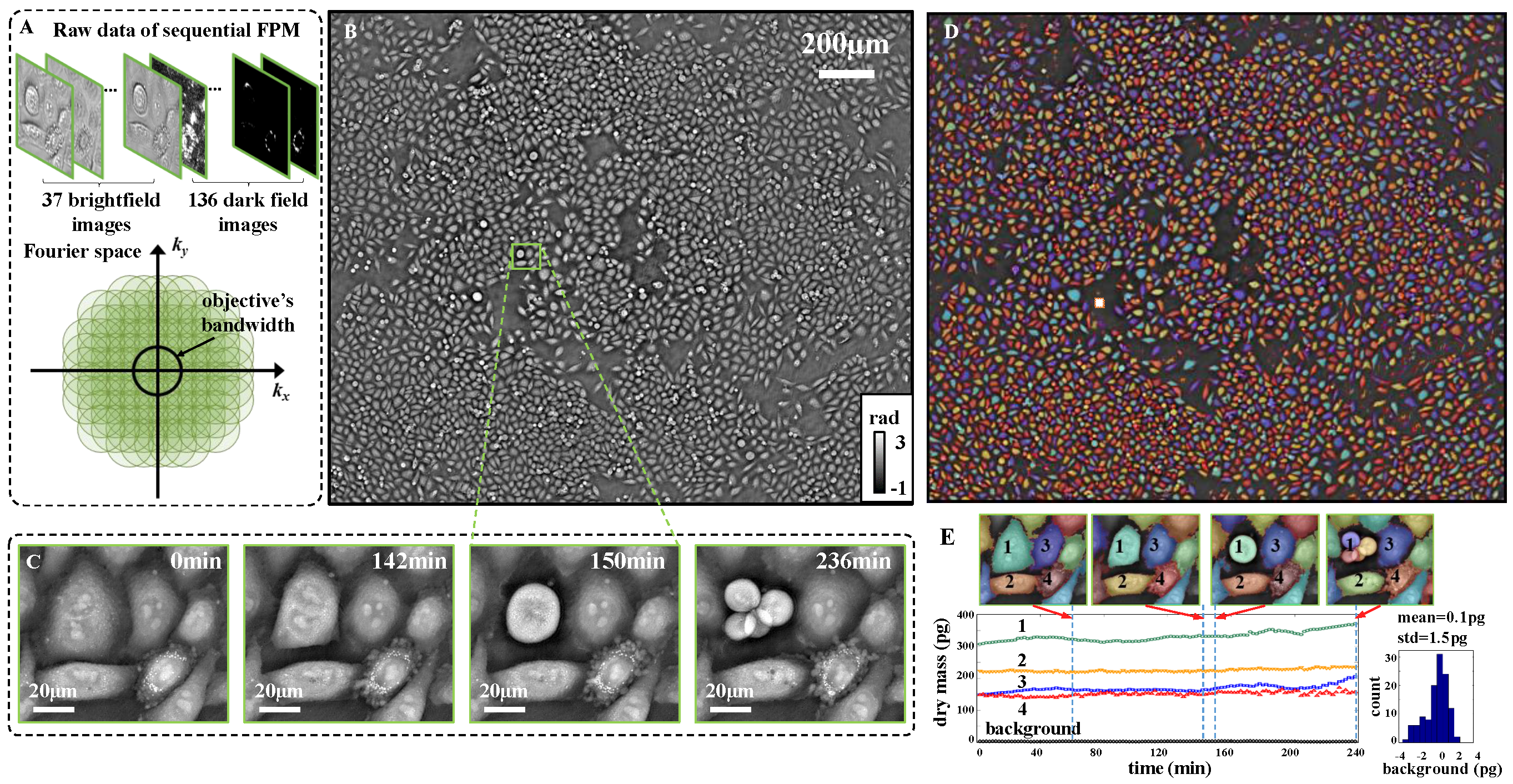}}
\caption{Time-lapse large SBP phase reconstruction of unstained HeLa cells undergoing division. (A) Sample raw data and Fourier coverage using sequential FPM (173 images), with acquisition time of 7 seconds per frame. (B) One frame of the full FOV phase reconstruction using a 4$\times$ objective and achieving 0.8 NA resolution. (C) Several frames of reconstructed video (see Video 1) from a zoom-in of one small area of confluent cells in which one cell is dividing into multiple cells. (D) Automated cell segmentation result for the full FOV phase image, with $\sim$3,400 cells successfully identified.
(E) Calculated dry mass for each of the labeled cells in the zoom-in region over 4 hours at 2 minutes intervals. At right is a histogram of the background fluctuations in an area with no cells.
\label{fig:Hela}}
\end{figure}

Figure~\ref{fig:Hela} shows a few frames from a time-lapse video of human cervical adenocarcinoma epithelial (HeLa) cell division process over the course of 4 hours, as well as an automated quantitative analysis of cell dry mass evolution. This data was captured using our improved sequential FPM (173 images) - sample raw images and a schematic of the Fourier space coverage are shown in Fig.~\ref{fig:Hela}A. One time frame of the full FOV phase reconstruction is shown in Fig.~\ref{fig:Hela}B, and a few frames of the video for a single zoom-in are shown in Fig.~\ref{fig:Hela}C. In this region, one cell is undergoing mitosis and dividing into 4 cells, during which the cells detach from the surface and become more globular (see Video 1). This situation, if imaged with a high magnification objective, would result in the detached cells moving out of the focal plane and blurring. However, FPM provides a longer DOF than a high magnification objective with equivalent NA, so the entire sample stays in focus across the FOV. Subcellular features are visible and the dynamics of actin filament formation can be tracked over time. We show only phase results (omitting intensity), since samples are unstained and have little intensity contrast.

Since our reconstructed video contains $\sim$20 gigapixels of quantitative phase data with $\sim$3,000 cells in each frame, the practical extraction of relevant information requires automated analysis. It is well known that PhC/DIC images cannot be automatically segmented (due to the lack of low spatial frequency information); however, our quantitative phase results do not have this problem. Using an automated cell segmentation software (Cellprofiler~\cite{Carpenter2006}) applied directly to the full FOV phase result, we first segment each frame of the video to find each of the $\sim$3,400 cells (Fig.~\ref{fig:Hela}D). Next, we compute each cell's dry mass~\cite{Popescu2008,Mir2011} over time through the division process, with a few sample cells plotted in Fig.~\ref{fig:Hela}E. Dry mass is calculated by integrating phase over each segmented cell region for each time frame (see Methods~\ref{method-segment}). Note that the automated cell segmentation and dry mass calculation are sensitive to the quality of the phase result, and often fail when low spatial frequencies are not well reconstructed; hence, the DPC initialization and other algorithm improvements implemented in this work are crucial for automated quantitative studies.

\subsection{Source-coded FPM video of neural stem cells \emph{in vitro}}  

\begin{figure}[!t]
\centerline{\includegraphics[width=0.75\textwidth]{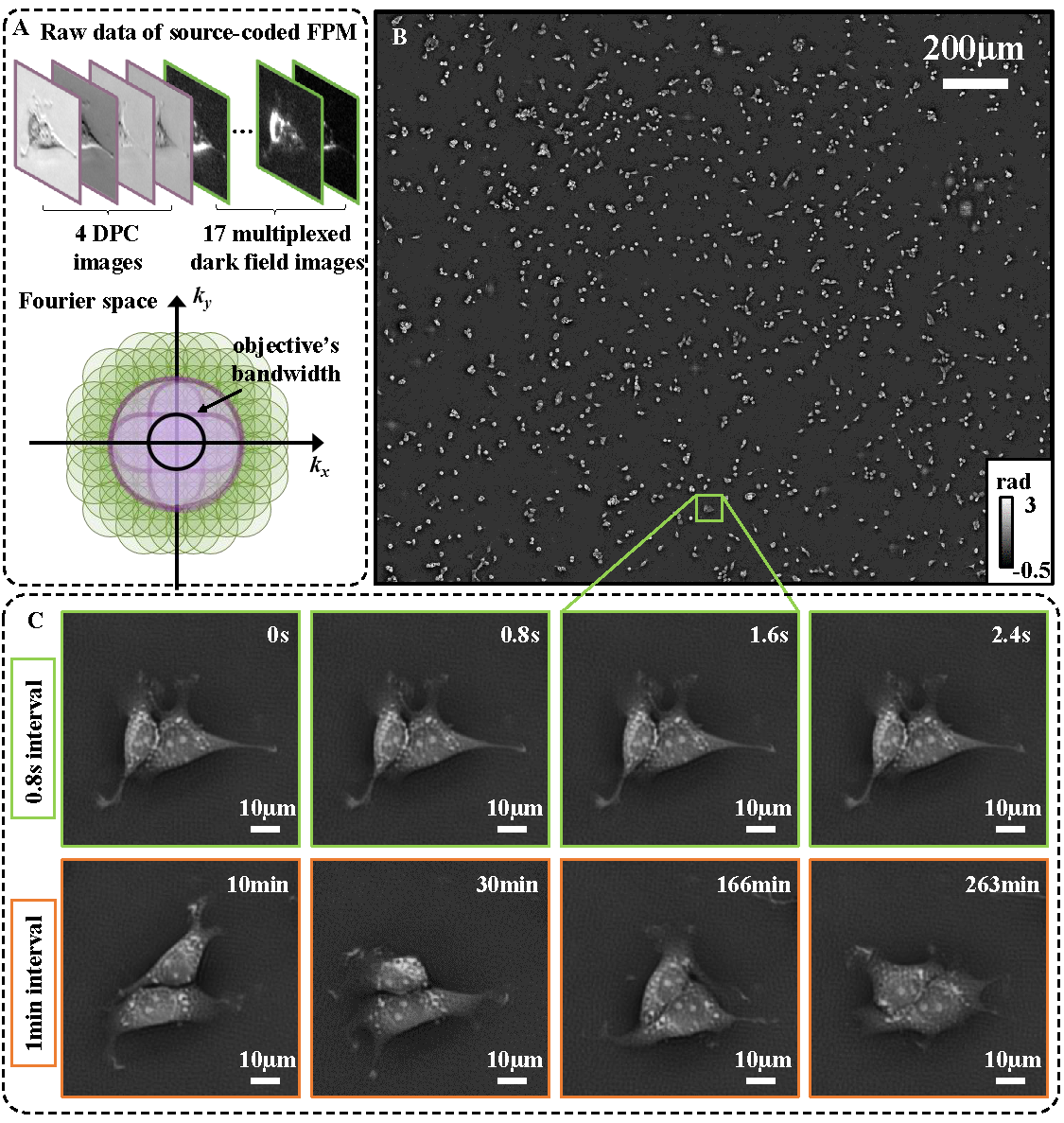}}
\caption{Large SBP phase video reconstructions for observing multi-scale temporal dynamics of \emph{in vitro} NSCs with high SBP and acquisition time of 0.8 seconds per frame. (A) Our source-coded FPM captures 4 brightfield images and 17 multiplexed dark field images. (B) Full FOV phase reconstruction using a 4$\times$ objective and achieving 0.8 NA resolution. (C) Sample frames of reconstructed video (see Video 2) for a zoom-in of one small area. Top: successive frames at the maximum frame rate (1.25 Hz), Bottom: sample frames across the longer time-lapse (4.5 hours at 1 minute intervals). 
\label{NSC}}
\end{figure}

Source-coded FPM can observe samples across long time scales (up to 4.5 hours) with sub-second acquisition speed (1.25 Hz). An example frame from a reconstructed large SBP phase video of adult rat neural stem cells (NSCs) is shown in Fig.~\ref{NSC}B, in a petri dish. We use source-coded FPM to achieve the same SBP as in sequential FPM (0.8 NA resolution across a 4$\times$ FOV), but with only 21 images (sample raw images shown in Fig.~\ref{NSC}A), as opposed to 173. As a result, we significantly decrease the capture time, from 7 seconds to 0.8 seconds. This alleviates motion artifacts that would otherwise blur the result due to sub-cellular dynamics that happen on timescales shorter than 7 seconds. Since our source-coded FPM reduces the number of images needed, we can also capture longer video sequences without incurring data management issues. Hence, we create videos of both fast-scale dynamics and slow-scale evolution of NSCs with details both on the sub-cellular level and across the entire cell population. For example, while vesicle transport and other organelle motions tend to occur on a short time scale and at a small length scale (see Video 2), NSC differentiation into neuronal and glial lineages occurs on a longer time scale (\emph{i.e.} days) and at a larger length scale. In the middle time scale, we observe NSCs sensing their micro-environment by retracting and extending processes, reorganizing their cytoskeletons, migrating, and maturing their axonal and dendritic processes.

\section{Discussion}

\begin{figure}[!t]
\centerline{\includegraphics[width=0.75\textwidth]{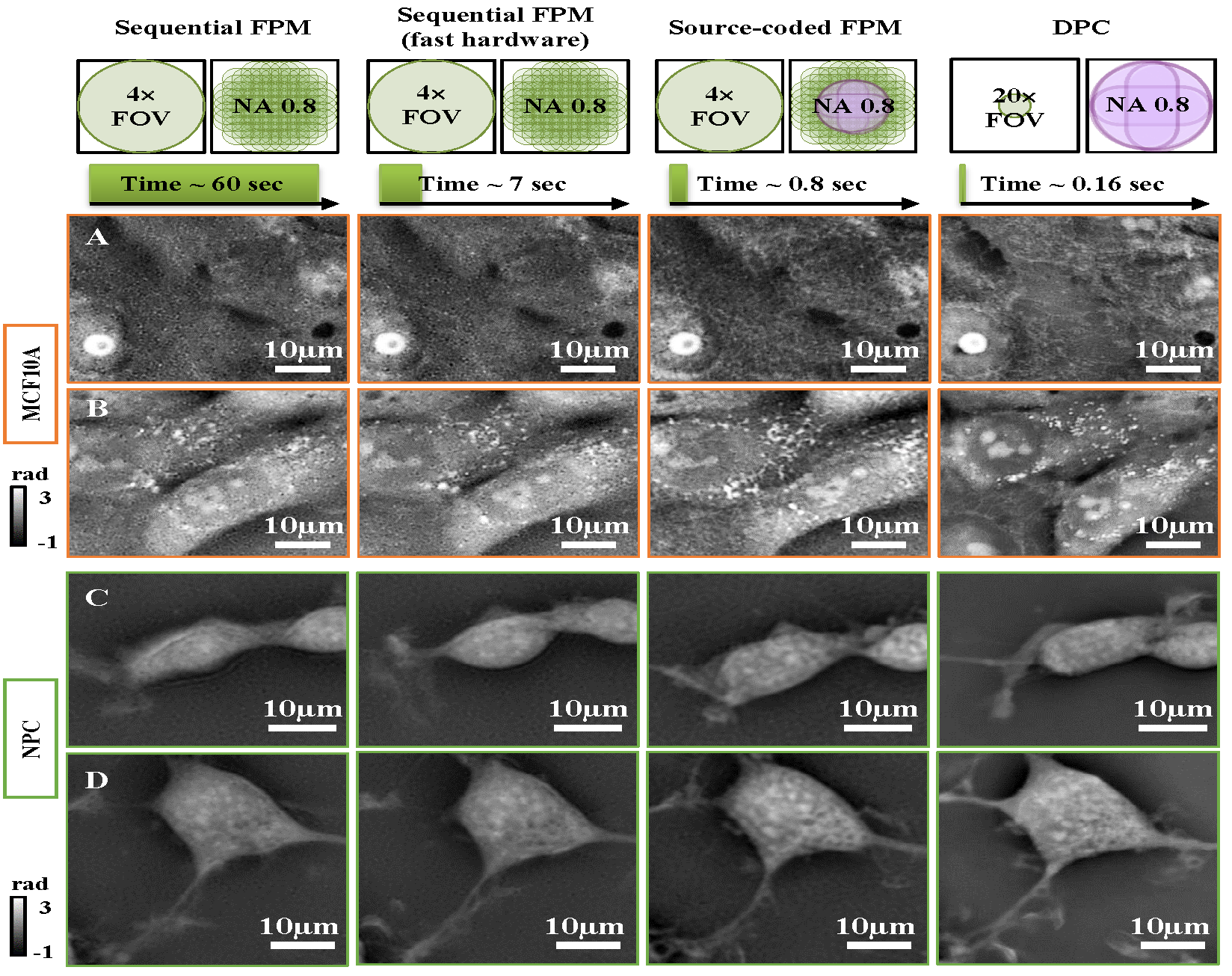}}
\caption{Motion blur degrades effective resolution in live dynamic samples. Reconstructed phase of live samples using different capture schemes with the same nominal spatial resolution (0.8 NA), but different acquisition times. As capture speed increases, more details about subcellular dynamics become visible due to reduced motion blur. Two fast dynamical processes in MCF10A cells: (A) subcellular fiber motion and (B) vesicle transport (Video 3) are blurred out when acquisition times are longer than 1 second. Our source-coded FPM achieves sub-second capture, revealing more details, yet not as clearly as DPC, which has the fastest capture time. (C) and (D) results for NSCs, which exhibit slower dynamics than the MCF10A cells. Sequential FPM blurs out most subcellular features; however, our source-coded FPM is able to capture details without motion artifacts (Video 4).
\label{fig:Motion}}
\end{figure}

With live samples imaged \emph{in vitro}, dynamics create motion blur artifacts that can destroy the resolution improvements gained by large SBP methods. Thus, the final effective resolution is always coupled with acquisition speed and sample-dependent motion.  In general, smaller subcellular features tend to move at faster speed; hence, we find that capture times on the order of minutes always incur motion blur. For this reason, faster acquisition is essential for \emph{in vitro} applications with typical cell types. To demonstrate this point, we compare the results of our source-coded FPM, sequential FPM with and without real-time hardware controls, and high magnification DPC. The DPC method is considered our benchmark, since it achieves faster acquisition speeds, avoiding motion blur, albeit with a small FOV.  In Fig.~\ref{fig:Motion}, we show zoomed-in phase reconstructions for two cell types with each capture scheme, from slowest to fastest. In each case, the final result has a nominal NA of 0.8 and so each of these results \emph{should} have the same resolution. However, it is clear that the slower capture schemes blur out features, rendering much of the small-scale structure and dynamics invisible. 

The MCF10A cells in Fig.~\ref{fig:Motion}A,B exhibit the fastest dynamics observed, due to rapid shuttling of small vesicles and fluctuations of cytoskeletal fibers (e.g. actin filaments, microtubules). Using sequential FPM with 60 second capture time, we retain almost no details about the structure of the fibers (Fig.~\ref{fig:Motion}A).  Even with our faster hardware setup, requiring only 7 seconds acquisition time, the fibers are still completely blurred out. By switching to our source-coded FPM, we obtain the same SBP as the previous two schemes, but with only 0.8 seconds acquisition time. Now, fiber cytoskeletal details become more discernible, though there is still some motion blur, as compared to the DPC result (Video 3). Hence, even our source-coded FPM is missing some information in this case. In Fig.~\ref{fig:Motion}B we zoom in on another fast process of vesicle transport, where we find that our method captures more of the dynamics than in the fiber fluctuation case, still with some blurring. In Video 3, we show time-lapse videos over 8 hours to observe cell migration and proliferation with subcellular details, captured using our source-coded FPM. Thus, the trade-off in FOV and time may be more appropriate, depending on what one aims to observe. 

In Fig.~\ref{fig:Motion}C,D, example results for two different NSCs are shown. In both cases, sequential FPM results in significant motion blur, particularly along thin, extended processes and within intracellular vesicles and organelles. In contrast, source-coded FPM is able to accurately capture the full details of the sample without motion blur artifacts. While it is difficult to compare directly because the live cells were moving in between capture schemes, in general we can say that most vesicle transport, retraction and extension of processes and other organelle motion can be clearly visualized using source-coded FPM (see Video 4).

The flexibility of our system in trading off FOV, resolution and time means that experiments can be tailored to the sample. For example, the HeLa and NSC samples shown here display slower sub-cellular dynamics, so are more suited to high SBP imaging with our current scheme, whereas the MCF10A sample necessitates a trade-off of SBP in order to capture data with sufficient speed. When dynamics are on timescales faster than 0.8 seconds, one should reduce the number of images captured, either by sacrificing FOV (using a larger NA objective) or by reducing the resolution improvement factor (using a smaller range of LEDs). In the limit, one can eliminate all the dark field LED images and simply implement DPC for maximum speed of capture. However, for most biological dynamics that are studied \emph{in vitro} (e.g. differentiation, division, apoptosis), we find that sub-second acquisition is sufficient, and so the SBP should be maximized within this constraint.

In summary, we have demonstrated a high-speed, large SBP microscopy technique, providing label-free quantitative phase and intensity information. Our source-coded FPM method overcomes the limitations of existing large SBP methods, permitting fast, motion-free imaging of unstained live samples. This work opens up large SBP imaging to high throughput \emph{in vitro} applications across a large range of both spatial and temporal scales. A gallery of interactive full FoV high resolution images from our experimental system can be found at
\url{http://www.gigapan.com/profiles/WallerLab_Berkeley}.

\section{Methods}
\subsection{Experimental setup}
\label{method-setup}
We place a custom-built 32$\times$32 surface-mounted LED array (4mm spacing, central wavelength 513nm with 20nm bandwidth) placed $\sim$70mm above the sample (Fig.~\ref{setup}A), replacing the microscope's standard illumination unit (Nikon TE300). All LEDs are driven statically using 64 LED controller chips (MBI5041) to provide independent drive channels. 
A controller unit based on ARM 32-bit Cortex\textsuperscript{TM} M3 CPU (STM32F103C8T6) provides the logical control for the LEDs by the I$^2$C interface at 5MHz, with LED pattern transfer time of $\sim$320$\mu$s. 
The camera (PCO.edge 5.5, 6.5$\mu$m pixel pitch) is synchronized with the LED array by the same controller via two coaxial cables which provide the trigger and monitor the exposure status. 
All raw images are captured with 14ms exposure times.  We experimentally measure the system frame rate to be $\sim$25Hz for capturing full-frame (2560 $\times$ 2160) 16 bit images. 
The data are transferred to the computer via CameraLink interface. All \emph{in vitro} experiments are performed in petri dishes placed inside a temperature and CO$_2$ controlled stage mounted incubator (In Vivo Scientific).

\subsection{Large SBP quantitative phase and intensity reconstruction}
\label{method-algorithm}
Our new FPM reconstruction algorithm can be described in two steps. First, we calculate a low-resolution initialization based on DPC. Next, we implement our quasi-Newton\rq{}s method iterative reconstruction procedure~\cite{Tian2014} to include the higher-order scattering and dark field contributions, for 3-5 iterations. In our source-coded FPM, the 4 brightfield images are directly used in the deconvolution-based DPC reconstruction algorithm~\cite{Tian2015a} to calculate phase within 2$\times$ the objective\rq{}s NA. The initial low-resolution intensity image is calculated by the average of all brightfield images corrected by the intensity falloffs~\cite{Phillips2015} and then deconvolved by the absorption transfer function~\cite{Tian2015a}. 
In our improved sequential FPM algorithm, the 4 DPC images are numerically constructed by taking the sum of single-LED images corresponding to the left, right, top, bottom half-circles on the LED array, respectively. 

In the reconstruction, we divided each full FOV raw image (2560$\times$2160 pixels) into 6$\times$5 sub-regions (560$\times$560 pixels each), with 160-pixel overlap on each side of neighboring sub-regions. Each set of images were then processed by our algorithm above to create a high resolution complex-valued reconstruction having both intensity and phase (2800$\times$2800 pixels). Finally, all high resolution reconstructions were combined using the alpha-blending stitching method~\cite{Zheng2013} to create the full FOV high resolution reconstruction. Using a desktop computer (Intel i7 CPU), the processing time for each sub-region was $\sim$30s in Matlab. The total processing time for each full FOV was $\sim$20min. 

\subsection{Phase reconstruction from through-focus intensity stack}
\label{method-throughfocus}
We capture intensity images with a high-resolution objective (40$\times$ 0.65 NA) while moving the sample axially to 17 exponentially spaced positions from -64$\mu$m to 64$\mu$m using a piezostage (MZS500-E-Z, Thorlabs). The images are then used to reconstruct phase (Fig.~\ref{compare}C) using a transport of intensity type algorithm based on spatial frequency domain fitting~\cite{Jingshan14GPTIE}.
 
\subsection{Simulation of phase contrast images}
\label{method-PhC}
Our phase contrast simulation (Fig.~\ref{compare}D) fully accounts for the partially coherent annular illumination (inner NA = 0.25, outer NA = 0.28, measured experimentally) and apodized phase contrast pupil~\cite{Otaki2000}. The pupil consists of a $\pi/2$-phase shifting ring with 75$\%$ attenuation (size matching the source), and two apodization rings with $50\%$ attenuation (widths are calculated based on~\cite{Otaki2000}). In our simulation, a tilted plane wave from each source point shifts the sample\rq{}s spectrum in the Fourier space, which is then filtered by the pupil function before calculating the intensity image in the real space. The phase contrast image is the incoherent sum of all the intensity contributions from all the points on the annular source~\cite{mandel1995optical}.

\subsection{Image segmentation and cell dry mass calculation}
\label{method-segment}
Image segmentation for each frame is performed by Cellprofiler~\cite{Carpenter2006} open-source software, which implements a series of automated operations including thresholding, watershedding and labeling, to return a 2D map containing segmented regions representing different cells. The 2D maps are then loaded into Matlab to extract the phase within each individual cell. The total dry mass for each cell is calculated as the sum of the dry mass density~\cite{Popescu2008}. The dry mass density $\rho$ is directly related to phase $\phi$ by $\rho=\frac{\lambda}{2\pi\gamma}\phi$, where $\lambda$ is the center wavelength, $\gamma=0.2$mL$/$g is the average of reported values for refractive increment of protein~\cite{Barer1952}. The background fluctuations are characterized from a region without any cells (white square in Figure~\ref{fig:Hela}D) and having an area similar to the average cell size. Background fluctuations are shown in the black curve and histogram in Figure~\ref{fig:Hela}E. We achieve a standard deviation of 1.5pg, in units of dry mass, indicating good stability of the phase measurement.

\subsection{Sample preparation}

HeLa cells were cultured with DMEM (Dulbecco’s Modified Eagle’s Medium) supplemented with 10\% FBS (fetal bovine serum), Glutamine and penicillin/streptomycin. The cells were plated on a p75 flask and cultured in a 37$^{\circ}$C incubator with 5\% CO$_2$ . The confluent cells were treated with 0.2\% trypsin and passed as 1:8 ratio. Two drops of trypisized cells were then added into a polyd-d-lysine coated 35mm MatTek glass-bottom plate with 2mL medium. After 6 hours incubation, the cells became fully attached to the plate. U2OS cells were prepared using the same procedure. They were fixed in 4\% formaldehyde at room temperature for 10 min and later stained with 1\% Toluidine Blue O (TBO) at room temperature for 5 min and rinsed in three changes of double-distilled water.
Adult rat NSCs were isolated from the hippocampi of 6-week-old female Fischer 344 rats. To promote adhesion, tissue culture polystyrene plates were first coated with 10$\mu$g/mL poly-L-ornithine (Sigma) overnight at room temperature, followed by 5$\mu$g/mL of laminin (Invitrogen) overnight at 37$^{\circ}$C. NSCs were cultured in monolayers in 1:1 DMEM/F12 high-glucose medium (Life Technologies), supplemented with N-2 (Life Technologies) and 20ng/mL recombinant human FGF-2 (Peprotech). Media was changed every other day, and cells were subcultured with Accutase upon reaching $\sim$80\% confluency. MCF10A cells were cultured in DMEM/F12 (Invitrogen), supplemented with 5\% horse serum (Invitrogen), 1\% penicillin/streptomycin (Invitrogen), 0.5$\mu$g/mL hydrocortisone (Sigma), 100 ng/mL cholera toxin (Sigma), 10$\mu$g/mL insulin (Sigma) and 20ng/mL recombinant human EGF (Peprotech). Media was changed every other day, and cells were passed with Trypsin upon reaching $\sim$80\% confluency. In preparation for imaging, cells were washed once with PBS (phosphate buffer saline) and detached with either Accutase or Trypsin. 300,000 cells were seeded onto 35 mm glass-bottom microwell dishes (MatTek) and allowed to attach.

\section*{Funding Information}
Funding was provided by the Gordon and Betty Moore Foundation's Data-Driven Discovery Initiative through Grant GBMF4562 to Laura Waller (UC Berkeley).
\section*{Acknowledgments}

We would like to thank Zachary Phillips for help with experiments, and Olivia Scheideler, Lydia Sohn, David Schaffer, Peiwu Qin and Ahmet Yildiz for providing cell samples and incubator.

\end{document}